\documentclass[%
 reprint,
superscriptaddress,
 amsmath,amssymb,
 aps,
prb,
]{revtex4-1}

\usepackage{graphicx}
\usepackage{dcolumn}
\usepackage{bm}

\begin{document}

\preprint{APS/123-QED}

\title{Linear dichroism in angle-resolved core-level photoemission spectra reflecting $4f$ ground-state symmetry of strongly correlated cubic Pr compounds}

\author{Satoru Hamamoto}
\author{Shuhei Fujioka}
\author{Yuina Kanai}
\author{Kohei Yamagami}
\author{Yasuhiro Nakatani}
\affiliation{Division of Materials Physics, Graduate School of Engineering Science, Osaka University, Toyonaka, Osaka 560-8531, Japan}
\affiliation{RIKEN SPring-8 Center, Sayo, Hyogo 679-5148, Japan}

\author{Koya Nakagawa}
\affiliation{RIKEN SPring-8 Center, Sayo, Hyogo 679-5148, Japan}
\affiliation{Faculty of Science and Engineering, Konan University, Kobe, Hyogo 658-8501, Japan}

\author{Hidenori Fujiwara}
\author{Takayuki Kiss}
\affiliation{Division of Materials Physics, Graduate School of Engineering Science, Osaka University, Toyonaka, Osaka 560-8531, Japan}
\affiliation{RIKEN SPring-8 Center, Sayo, Hyogo 679-5148, Japan}

\author{Atsushi Higashiya}
\affiliation{Faculty of Science and Engineering, Setsunan University, Neyagawa, Osaka 572-8508, Japan }
\affiliation{RIKEN SPring-8 Center, Sayo, Hyogo 679-5148, Japan}

\author{Atsushi Yamasaki}
\affiliation{RIKEN SPring-8 Center, Sayo, Hyogo 679-5148, Japan}
\affiliation{Faculty of Science and Engineering, Konan University, Kobe, Hyogo 658-8501, Japan}

\author{Toshiharu Kadono}
\author{Shin Imada}
\affiliation{RIKEN SPring-8 Center, Sayo, Hyogo 679-5148, Japan}
\affiliation{Department of Physical Science, Ritsumeikan University, Kusatsu, Shiga 525-8577, Japan }

\author{Arata Tanaka}
\affiliation{Department of Quantum Matter, ADSM, Hiroshima University, Higashi-Hiroshima, Hiroshima 739-8530, Japan}

\author{Kenji Tamasaku}
\author{Makina Yabashi} 
\author{Tetsuya Ishikawa}
\affiliation{RIKEN SPring-8 Center, Sayo, Hyogo 679-5148, Japan}

\author{Keisuke T. Matsumoto}
 \altaffiliation[Present address: ]{Department of Materials Science and Biotechnology, Ehime University, Matsuyama, Ehime 790-8577, Japan}
\author{Takahiro Onimaru}
\affiliation{Department of Quantum Matter, ADSM, Hiroshima University, Higashi-Hiroshima, Hiroshima 739-8530, Japan}

\author{Toshiro Takabatake}
\affiliation{Department of Quantum Matter, ADSM, Hiroshima University, Higashi-Hiroshima, Hiroshima 739-8530, Japan}
\affiliation{Institute for Advanced Materials Research, Hiroshima University, Higashihiroshima, Hiroshima 739-8530, Japan}

\author{Akira Sekiyama}
\affiliation{Division of Materials Physics, Graduate School of Engineering Science, Osaka University, Toyonaka, Osaka 560-8531, Japan}
\affiliation{RIKEN SPring-8 Center, Sayo, Hyogo 679-5148, Japan}

\date{\today}

\begin{abstract}
We report experimentally observed linear dichroism in angle-resolved core-level photoemission spectra of PrIr$_2$Zn$_{20}$ and PrB$_6$ in cubic symmetry. The different anisotropic $4f$ charge distributions between the compounds due to the crystalline-electric-field splitting are responsible for the difference in the linear dichroism, which has been verified by spectral simulations with the full multiplet theory for a single-site Pr$^{3+}$ ion in cubic symmetry. The observed linear dichroism and polarization-dependent spectra in two different photoelectron directions for PrIr$_2$Zn$_{20}$ are reproduced by theoretical analysis for the $\Gamma_3$ ground state, whereas those of the Pr $3d$ and $4d$ core levels indicate the $\Gamma_5$ ground state for PrB$_6$.
\end{abstract}

\maketitle

\section{Introduction}
Rare earth compounds have been investigated for many years in relation to the physical properties of the strongly correlated electron systems, such as heavy fermion behavior, unconventional superconductivity, and multipole ordering.
In these systems, the total angular momentum $J$ of the $4f$ electrons is the good quantum number due to the stronger Coulomb interactions between the $4f$ electrons and their spin-orbital coupling than the crystalline electric field (CEF). Therefore, the observable is not the orbital degrees of freedom but the quadrupoles which are described as second-order tensors of $J$. When the dipole magnetic moments are quenched but the quadrupoles remain active under CEF, which is possible in non-Kramers ions with an integer number of $J$ under cubic symmetry, the quadrupoles often play an important role in forming exotic electronic ground states such as the antiferroquadrupolar (AFQ) state with a staggered quadrupolar component,\cite{morin} and a non-Fermi liquid (NFL) state attributed to the two-channel (quadrupole) Kondo effect.\cite{cox1,cox2,onimaru0} Furthermore, the feasibility of superconductivity mediated by quadrupolar fluctuations in the Pr-based superconductor PrOs$_4$Sb$_{12}$ with $T_c$ = 1.5 K\cite{bauer} has been pointed out by neutron scattering and NQR measurements.\cite{kuwahara,yogi} A well-known example of a $4f^2$ cubic system is PrPb$_3$ which undergoes an AFQ transition at $T_Q$ = 0.4 K.\cite{tayama,onimaru00}

Here reported cubic PrIr$_2$Zn$_{20}$ undergoes an AFQ ordering at $T_Q$ = 0.11 K and a superconducting transition at $T_c$ = 0.05 K.\cite{onimaru1,onimaru2} Below 10 K, the increase in magnetic susceptibility for this compound tends to saturate, which implies the van Vleck susceptibility with nonmagnetic CEF ground states of the $\Gamma_3$ doublet.\cite{onimaru3} On the other hand, PrB$_6$ with a cubic crystal structure show the incommensurate antiferromagnetic transition at 7 K and a commensurate antiferromagnetic ordering at 4.2 K.\cite{kobayashi} These phenomena suggest the $\Gamma_5$ ground state for PrB$_6$. However, direct or microscopic verifications of the proposed $4f$ ground-state symmetry are still lacking for these compounds. In this letter, we show spectroscopic signs for the $\Gamma_3$ symmetry of the Pr$^{3+}$ sites in the ground state of PrIr$_2$Zn$_{20}$ and the $\Gamma_5$ ground state of PrB$_6$ on the basis of linear dichroism in angle-resolved core-level photoemission from the Pr sites.

Experimentally, it is difficult to detect the anisotropic $4f$ charge distributions in cubic symmetry. 
Although inelastic neutron scattering is useful for revealing the CEF splittings, 
the symmetry itself cannot be directly probed by unpolarized neutrons. 
Linear dichroism (LD) in $3d$-$4f$ soft X-ray absorption spectroscopy (XAS) for single crystals is powerful for tetragonal Ce compounds\cite{hansman,willers1,willers2,willers3} owing to the dipole selection rules, but it is no longer applicable for compounds with cubic crystal structure.
On the other hand, the selection rules work also in the photoemission process. Furthermore, by virtue of the angle-resolved measurements with the acceptance angle of $\pm$several degrees, there is another measurement parameter as the photoelectron direction relative to the single-crystalline axis in addition to the excitation light polarization. Indeed, by using LD in $3d$ core-level hard X-ray photoemission (HAXPES) spectra, the Yb$^{3+}$ $4f$ ground state has been determined for 
cubic YbB$_{12}$.\cite{kanai} LD in the core-level HAXPES for cubic Pr compounds is also expected to be observed, as discussed below.

\begin{figure}[t]
\begin{center}
\includegraphics[width=6cm,clip]{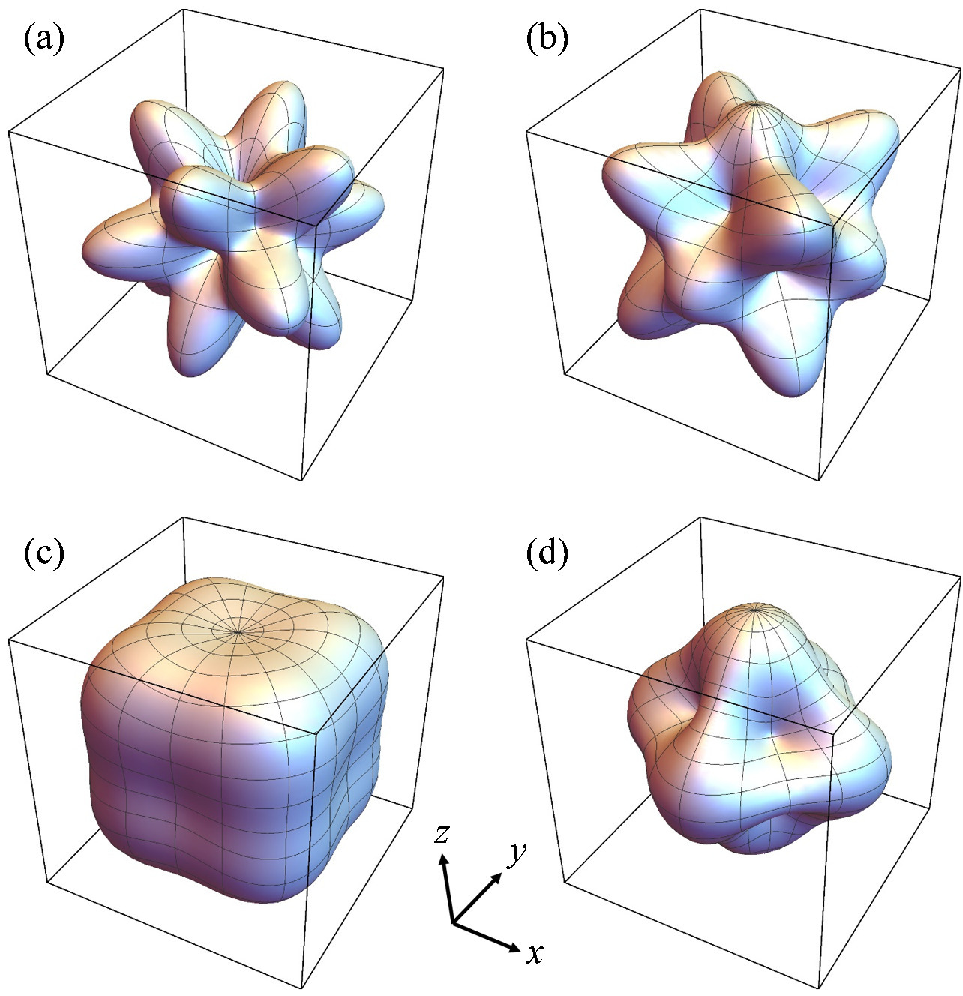}
\caption{(Color online) $4f^2$ charge distributions for the (a) $\Gamma_1$, (b) $\Gamma_3$, (c) $\Gamma_4$, and (d) $\Gamma_5$ states, respectively, obtained from Eqs. (\ref{wavefunction1} - \ref{wavefunction5}).}
\label{chargedistribution}
\end{center}
\end{figure}

The Pr ions are encapsulated in the highly symmetric Frank-Kasper cages formed by 16 zinc atoms in PrIr$_2$Zn$_{20}$, where the actual local point-group symmetry is $T_d$.\cite{nasch} For PrB$_6$, on the other hand, the Pr ions are in the $O_h$ symmetry.\cite{maccarthy} Because of the even number of $4f$ electrons, the CEF electronic states are free from the Kramers theorem. The ninefold degenerated ground-state multiplets with $J$ = 4 ($L$ = 5, $S$ = 1) in the $4f^2$ configurations determined by the Hund's rules is lifted by the CEF Hamiltonian for the $T_d$ or $O_h$ symmetry,
\begin{align}
H_{CEF}=B_{40} (O_{40}+5O_{44})+B_{60} (O_{60}-21O_{64}),
\label{hamiltonian}
\end{align}
where $O_{mn}$ and $B_{mn}$ denote the Stevens operator and the CEF parameter, respectively.\cite{stevens} 
It has been reported that candidates of the CEF parameters ($B_{40}$, $B_{60}$) are ($-1.1 \times 10^{-2}$ K, $-4.5 \times 10^{-4}$ K)\cite{iwasa} and ($1.5 \times 10^{-1}$ K, $3.8 \times 10^{-4}$ K)\cite{loewenhaupt} for PrIr$_2$Zn$_{20}$ and PrB$_6$, respectively. 
Then the ninefold $J$ = 4 states split generally into one singlet, one doublet, and two triplets as
\begin{align}
&|{\Gamma_1}\rangle=\sqrt{5/24}|{+4}\rangle+\sqrt{7/12}|{0}\rangle+\sqrt{5/24}|{-4}\rangle ,& \label{wavefunction1} \\
&|{\Gamma_3}\rangle=\left\{
\renewcommand{\arraystretch}{1.5}
\begin{array}{l}
\sqrt{7/24}|{+4}\rangle-\sqrt{5/12}|{0}\rangle+\sqrt{7/24}|{-4}\rangle \\
\sqrt{1/2}|{+2}\rangle+\sqrt{1/2}|{-2}\rangle \label{wavefunction3} \\
\end{array}
\renewcommand{\arraystretch}{1}
,\right.& \\
&|{\Gamma_4}\rangle=\left\{
\renewcommand{\arraystretch}{1.5}
\begin{array}{l}
\sqrt{1/8}|{\pm 3}\rangle+\sqrt{7/8}|{\mp 1}\rangle \\
\sqrt{1/2}|{+4}\rangle-\sqrt{1/2}|{-4}\rangle \label{wavefunction4} \\
\end{array}
\renewcommand{\arraystretch}{1}
,\right.& \\
&|{\Gamma_5}\rangle=\left\{
\renewcommand{\arraystretch}{1.5}
\begin{array}{l}
\sqrt{7/8}|{\pm 3}\rangle-\sqrt{1/8}|{\mp 1}\rangle \\
\sqrt{1/2}|{+2}\rangle-\sqrt{1/2}|{-2}\rangle \label{wavefunction5} \\
\end{array}
\renewcommand{\arraystretch}{1}
,\right.&
\end{align}
where the integers in the kets denote $J_z$. It is known that the $\Gamma_4$ states are ruled out from the candidates of the ground state because there is no solution of $B_{40}$ and $B_{60}$ to set them as the lowest state.\cite{llw} Since their $4f$ charge distributions deviate from spherical symmetry owing to the CEF splitting even in the cubic symmetry, as shown in Fig. \ref{chargedistribution}, it is natural to expect the observation of LD in core-level photoemission for cubic Pr compounds. Actually, we have performed ionic calculations including the full multiplet theory\cite{thole} and the local CEF splitting using the XTLS 9.0 program,\cite{tanaka} by which finite LDs as a function of the photoelectron direction have been predicted as discussed below. The atomic parameters such as the $3d$-$4f$ Coulomb and exchange interactions (Slater integrals) and the $3d$ and $4f$ spin-orbit couplings have been obtained using Cowan's code\cite{cowan} based on the Hartree-Fock method. The $4f$-$4f$ Slater integrals have been obtained using Cowan's code and their relations.\cite{carnall} The $3d$-$4f$ and $4f$-$4f$ Slater integrals ($3d$ and $4f$ spin-orbit couplings) are reduced to 80\% and 60\% (99\% and 100\%) to reproduce the Pr$^{3+}$ $3d$ core-level photoemission spectra with final-state multiplet structures including Pr$^{3+}$ $3d_{3/2}$ spectra, respectively.

\section{Experimental}
We have observed LD in HAXPES\cite{haxpes2,mori,fujiwara} at BL19LXU of SPring-8\cite{bl19lxu} using a MBS A1-HE hemispherical photoelectron spectrometer. A Si(111) double-crystal monochromator selected linearly polarized 7.9 keV radiation within the horizontal plane, which was further monochromatized using a Si(620) channel-cut crystal. To switch the linear polarization of the excitation light from the horizontal to vertical directions, two single-crystalline (100) diamonds were used as a phase retarder placed downstream of the channel-cut crystal. The $P_L$ (degree of linear polarization) of the polarization-switched X-ray after the phase retarder was estimated as $-0.89$, corresponding to the vertically linear polarization component of 93.8\%. Since the detection direction of photoelectrons was set in the horizontal plane with an angle to incident photons of 60$^\circ$, the experimental configuration at the vertically (horizontally) polarized light excitation corresponds to the s-polarization (p-polarization). To precisely detect LD in the Pr $3d$ core-level photoemission spectra, we optimized the photon flux so as to set comparable photoelectron count rates between the s- and p-polarization configurations. Single crystals of PrIr$_2$Zn$_{20}$ synthesized by the melt-growth method\cite{onimaru2,saiga} were fractured along the (110) plane {\it in situ}, where the base pressure was $1 \times 10^{-7}$ Pa. The experimental geometry was controlled using our developed two-axis manipulator,\cite{fujiwara} where the photoelectron detection along the [100] direction was set by polar rotation of 45$^\circ$ from the normal direction parallel to the [110] direction, and the [111] direction was set by polar rotation by $\sim$35$^\circ$ after azimuthal rotation. A single crystal of PrB$_6$ synthesized by the floating zone method was fractured along the (100) plane {\it in situ}. The sample and surface qualities were examined on the basis of the absence of O and C $1s$ core-level spectral weights caused by possible impurities or surface oxidization. The energy resolution was set to 500 meV. The measuring temperature is 5 K (10 K) for PrIr$_2$Zn$_{20}$ (PrB$_6$), which is sufficiently lower than the excited states.\cite{iwasa}

\section{Results and Discussions}

\begin{figure}[h]
\begin{center}
\includegraphics[width=8.5cm,clip]{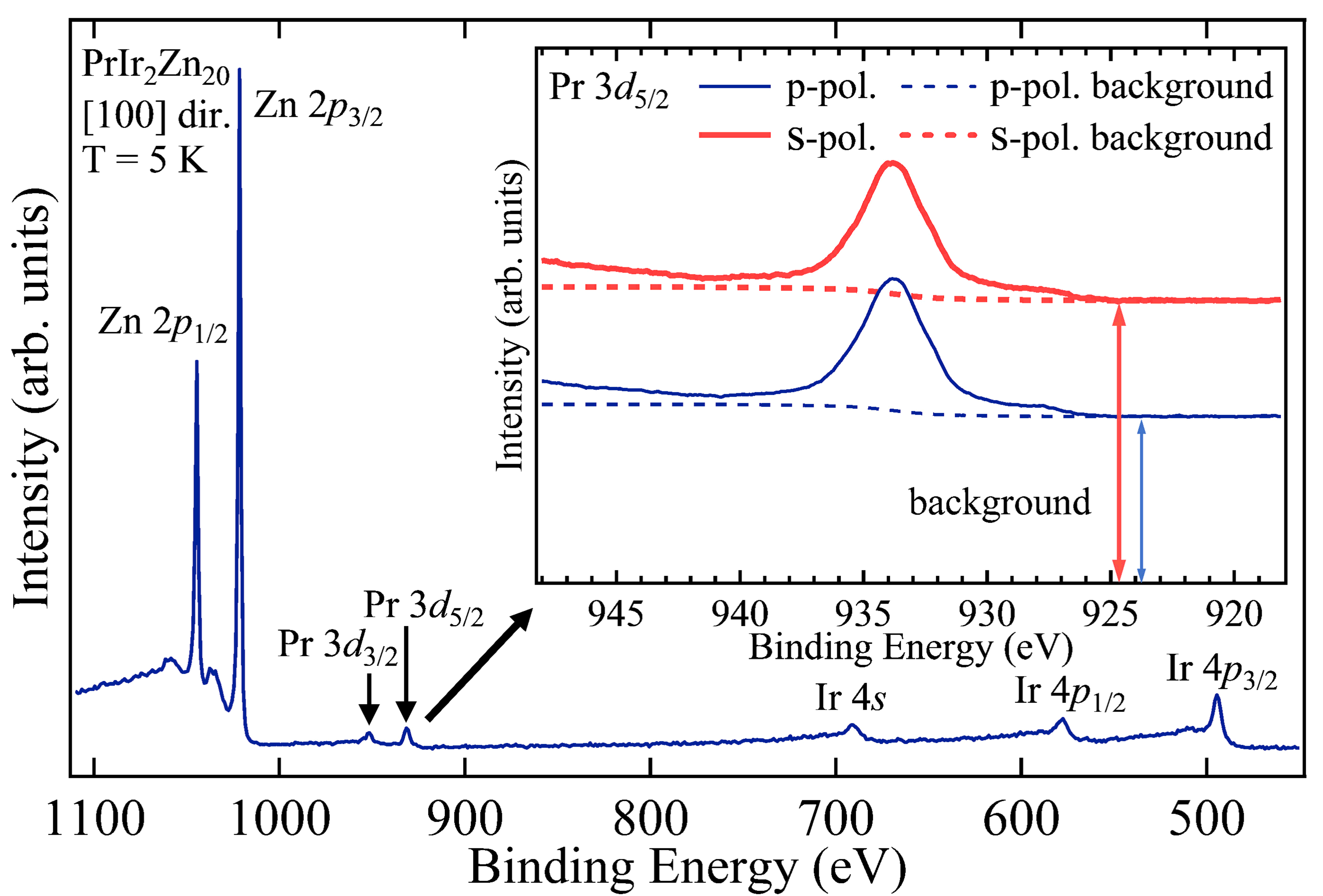}
\caption{(Color online) Core-level HAXPES raw spectra in the p-polarization configurations. The inset shows polarization-dependent Pr $3d_{5/2}$ core-level HAXPES raw spectra (solid lines) of PrIr$_2$Zn$_{20}$ in the [100] direction and optimized Shirley-type backgrounds (see text), which we have subtracted from the raw spectra (dashed lines).}
\label{pr3d_bg}
\end{center}
\end{figure}

The polarization-dependent Pr$^{3+}$ $3d_{5/2}$ HAXPES spectra in the [100] direction of PrIr$_2$Zn$_{20}$ are shown in Fig. \ref{pr3d_bg}. Caused by the relatively low density of the Pr ions as $<$5\% determined by the composition, the intrinsic Pr $3d_{5/2}$ signals are much weaker than the background contributed by the photoelectrons from the other orbitals in the other elements. A broad peak at a binding energy of $\sim$934 eV, shoulder centered at 939 eV and another shoulder structure ranging from 925 to 930 eV exist in the spectra. The former peak and the shoulder at 939 eV are ascribed to the Pr$^{3+}$ states ($|{3d^94f^2}\rangle$ final states) with the atomic-like multiplets structure, and the shoulder structure at $925-930$ eV comes from the $|{3d^94f^3}\rangle$ final states with hybridization between the $4f$ and valence/conduction electrons. The so-called Shirley-type backgrounds are also displayed in Fig. \ref{pr3d_bg}.
We have optimized the backgrounds as follows: After the normalization of the background-subtracted spectra by 
the $3d_{5/2}$ spectral weight in the $931-940$ eV region corresponding to 
the $|3d^94f^2\rangle$ final-state (ionic Pr$^{3+}$) region,
the intensities in the high binding energy region of $941-942$ eV become equivalent between the s- and p-polarization configurations. When we have normalized the spectra by the $3d_{5/2}$ spectral weight in the region included both the $|{3d^94f^2}\rangle$ and $|{3d^94f^3}\rangle$ final states, the LD is slightly suppressed but the sign of the LD is not changed. Therefore, we discuss below by the normalized spectra by only the spectral weight in the $|{3d^94f^2}\rangle$ final states.
The reference binding energy on the higher side for subtracting the background has been set to 940.8 eV corresponding to the local minimum of the raw spectral weight. 
As a result, there are finite spectral weights at $\sim$940 eV in the background-subtracted spectra. 
However, they should be intrinsic due to the overlap of the tails of the lifetime-broadened Pr$^{3+}$ $3d_{5/2}$ main peaks and a plasmonic energy-loss structure at the higher binding energy. 

\begin{figure}[h]
\begin{center}
\includegraphics[width=8cm,clip]{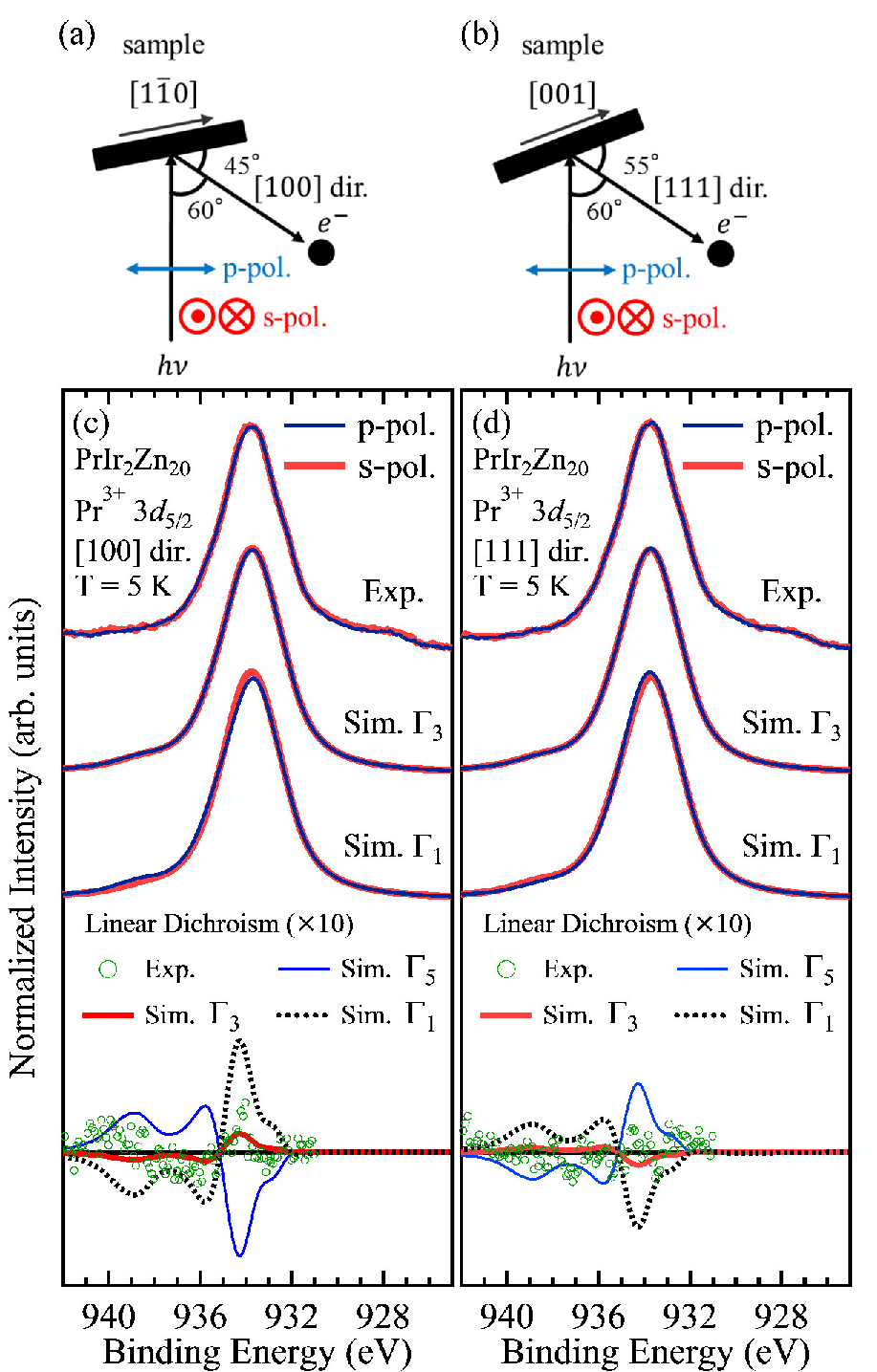}
\caption{(Color online) (a) Geometry for the polarization-dependent HAXPES measurements in the [100] direction for PrIr$_2$Zn$_{20}$. (b) same as (a) but for the photoelectron in the [111]
direction. (c) Polarization-dependent Pr $3d_{5/2}$ core-level HAXPES spectra and LD of PrIr$_2$Zn$_{20}$ compared with the simulated ones for the $\Gamma_3$ and $\Gamma_1$ states in the [100] direction. The experimental LD is displayed in the energy region of the normalization ($931-940$ eV). Simulated LD assuming the $\Gamma_5$ state is also shown in the lower panel. The Shirley-type background has been subtracted
from the raw spectra (Fig. \ref{pr3d_bg}). (d) Same as (c) but data in the [111] direction.}
\label{pr1220_ld}
\end{center}
\end{figure}

Figure \ref{pr1220_ld} shows comparisons of the polarization-dependent background-subtracted Pr$^{3+}$ $3d_{5/2}$ HAXPES spectra of PrIr$_2$Zn$_{20}$ and their LD (defined as the difference in the spectral weight between the s- and p-polarization configurations) with the photoelectron directions of [100] and [111] with the simulated ones for the $\Gamma_3$ and $\Gamma_1$ states. 
Here, we have simulated the spectra for the $\Gamma_3$ state at 5 K using the CEF parameters for PrIr$_2$Zn$_{20}$ listed above, which gives the first excited CEF level of 20 K.
For the experimental spectra and the simulated spectra for the $\Gamma_3$ state, the LDs are found to be finite but subtle. As shown in Fig. \ref{pr1220_ld}(c), the sign of the experimental LD at $\sim$934 eV is positive, whereas that in the $935-938$ eV region is negative along the [100] direction. These tendencies are consistent with the simulated LDs for the $\Gamma_3$ and  $\Gamma_1$ states although the LD for the $\Gamma_1$ state is much stronger at $\sim934$ eV than the experimentally observed LD. For the data in the [111] direction in Fig. \ref{pr1220_ld}(d), the LD in the $932-940$ eV region in the experiment is strongly suppressed compared with that in the [100] direction, being consistent with, at least not contradictory to, the simulation for the $\Gamma_3$ state. If the $4f$ ground state of PrIr$_2$Zn$_{20}$ were in the $\Gamma_1$ symmetry, a finite LD beyond the statistics would be seen with the negative sign at $\sim934$ eV. The sign and amount of LD for the $\Gamma_5$ state are completely inconsistent with our experimental results in both directions. Therefore, we can conclude that our experimental results support the $\Gamma_3$ ground-state symmetry for PrIr$_2$Zn$_{20}$. 
The LDs are positive in the $|{3d^94f^3}\rangle$ final-state region with the binding energies of 928-930 eV for both directions, for which the origin is not clear.
The relatively smaller LDs in the [111] direction than in the [100] direction in both experiments and simulations originate from the larger number of the equivalent direction (e.g. [100], [010] and [001]) in the cubic unit cell (8 for [111] and 6 for [100]), where the simulated LD in the [110] direction (not shown here) is much smaller than that in the [100] and [111] directions. (Note that the LD cancels out for really angle-integrated spectra over 4$\pi$ steradian.)

\begin{figure}[t]
\begin{center}
\includegraphics[width=8cm,clip]{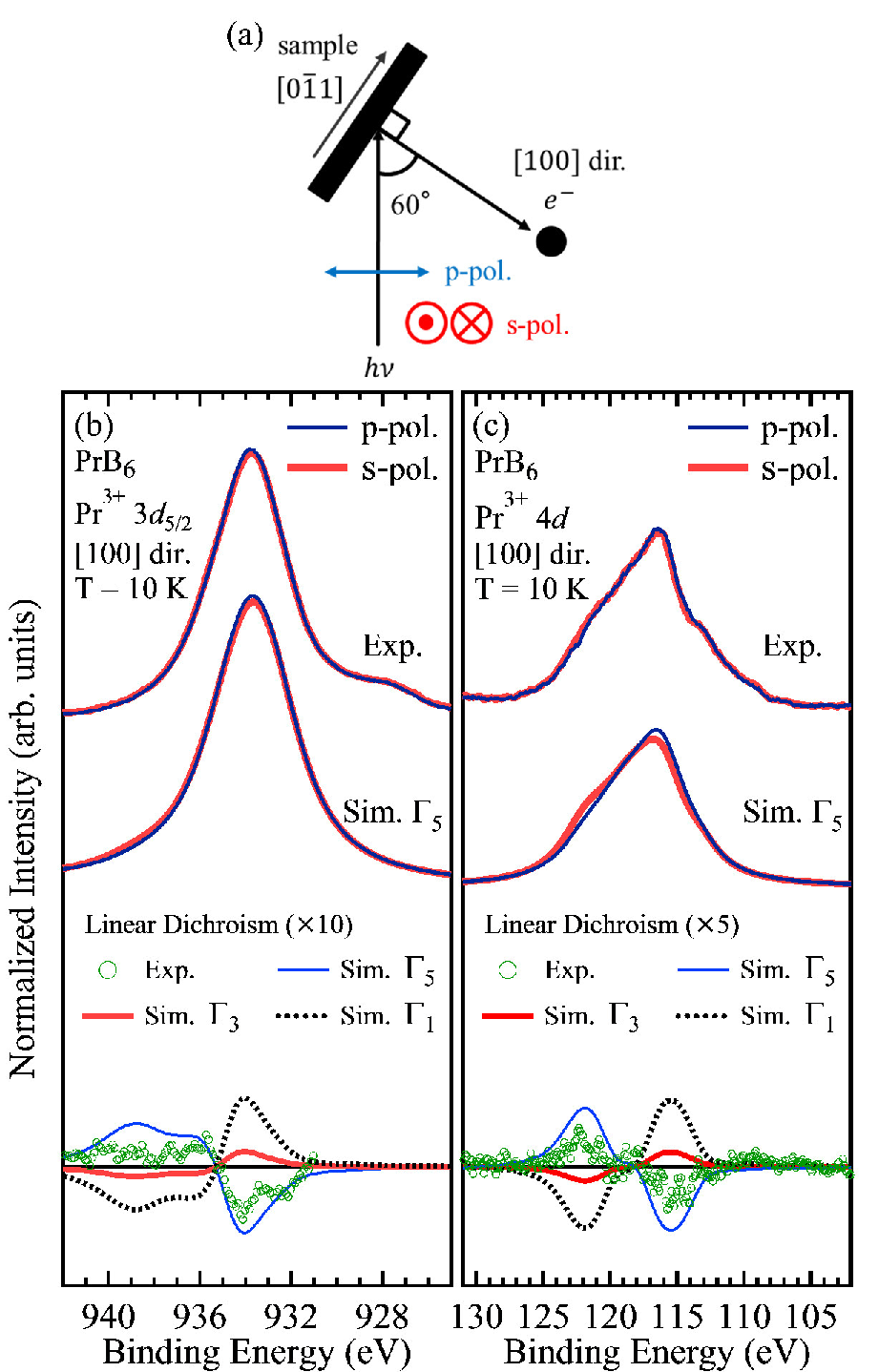}
\caption{(Color online) (a) Geometry for the polarization-dependent HAXPES  in the [100] direction for PrB$_6$. (b) Polarization-dependent Pr $3d_{5/2}$ core-level HAXPES spectra and LD of PrB$_6$ compared with the simulated ones for the $\Gamma_5$ state in the [100] direction as a manner of Figs. \ref{pr1220_ld}. Simulated LDs assuming the $\Gamma_3$ and $\Gamma_1$ states are also shown in the lower panel. (c) Same as (b) but for the Pr $4d$ core-level HAXPES.}
\label{prb6_ld}
\end{center}
\end{figure}

Figure \ref{prb6_ld} shows the polarization-dependent background-subtracted Pr$^{3+}$ $3d_{5/2}$ and $4d$ core-level HAXPES spectra and LDs of cubic PrB$_6$ in the [100] direction. In the case of PrB$_6$, the broad peak and the shoulder structures are seen in the Pr$^{3+}$ $3d_{5/2}$ spectra. On the other hand, there are multiplets ranging from 114 to 127 eV and a shoulder structure ranging from 106 to 114 eV in the Pr$^{3+}$ $4d$ spectra. The former multiplets is ascribed to the Pr$^{3+}$ states ($|{4d^94f^2}\rangle$ final states), and the latter shoulder structure comes from the $|{4d^94f^3}\rangle$ final states with hybridization between the $4f$ and conduction electrons. These features are qualitatively consistent with the previous reported ones.\cite{suga} The $3d_{5/2}$ main peak is broader for PrB$_6$ than for PrIr$_2$Zn$_{20}$ in Fig. \ref{pr1220_ld}(c), of which the origin is unclear at present.
It is difficult to distinguish clearly the structure of the $|{4d^94f^2}\rangle$ final states from that of the $|{4d^94f^3}\rangle$ final states in the Pr $4d$ spectra. Therefore, Pr $4d$ spectra have been normalized by spectral weight in the region included both the $|{4d^94f^2}\rangle$ and $|{4d^94f^3}\rangle$ final states.

The simulated polarization-dependent Pr$^{3+}$ $3d_{5/2}$ ($4d$) HAXPES spectra for the $\Gamma_5$ state 
at 10 K with the CEF parameters for PrB$_6$ listed above, which gives the first excited CEF level of 300 K,
and LDs for the $\Gamma_1$, $\Gamma_3$, and $\Gamma_5$ states are also shown in Fig. \ref{prb6_ld}. For obtaining the simulated Pr$^{3+}$ $4d$ spectra, the $4d$-$4f$ Slater integrals ($4d$ spin-orbit coupling) are reduced to 80\% (100\%) to best reproduce the experimental Pr$^{3+}$ $4d$ core-level photoemission spectra with respect to the multiplet splittings. To reproduce the spectra, the Lorentian width of 3.0 eV (full width of half maximum), which is larger than the value for PrIr$_2$Zn$_{20}$ of 1.8 eV as shown in Fig. \ref{pr1220_ld}(c), has been employed for PrB$_6$. The peak at 934 eV is slightly stronger in the p-polarization configuration than in the s-polarization one in the experimental Pr$^{3+}$ $3d_{5/2}$ spectra along the [100] direction, which is in contrast to the case of PrIr$_2$Zn$_{20}$. This tendency and LD are reproduced by the simulations for the $\Gamma_5$ states as shown in Fig. \ref{prb6_ld}(a). In the case of the Pr$^{3+}$ $4d$ spectra, the highest peak at a binding energy of $\sim$116 eV is slightly stronger in the p-polarization configuration than in the s-polarization one for both experimental and simulated spectra while the sign of LD is flipped in the binding energies of $120-125$ eV. As shown in Figs. \ref{prb6_ld}(b) and \ref{prb6_ld}(c), the observed LDs are qualitatively reproduced by the simulations for the $\Gamma_5$ state although LDs are somehow smaller in the experiment. 
Furthermore, the LD signs for the $\Gamma_3$ and $\Gamma_1$ states are completely opposite to those of the experimental data in all binding energy region. Thus, our results undoubtedly indicate the $\Gamma_5$ ground-state symmetry for PrB$_6$, being consistent with the previous report.\cite{loewenhaupt} 
The origin for the reduced LD in the experiment compared with the simulations for the Pr$^{3+}$ ion in cubic symmetry is not clear at present, but we can speculate a couple of possible origins. One is the $c$-$f$ hybridization effects giving the $|3d^94f^3\rangle$ final states. The other is a possible antiferromagnetic fluctuation at 10 K just above the transition temperature of 7 K (Ref. \cite{kobayashi}). Further detailed temperature and angular dependence of the LD could solve the problem in future.
Note that the Pr $4d$ spectral weight has not unfortunately been detected for PrIr$_2$Zn$_{20}$ since it has been buried in the background from the other sites by the HAXPES at $h\nu$ = 7.9 keV.

\section{Summary}
In summary, we have performed the polarization-dependent core-level HAXPES of the Pr$^{3+}$ sites in cubic PrIr$_2$Zn$_{20}$ and PrB$_6$. Our results support the so far predicted their ground-state symmetry, showing the potential of LD in the core-level HAXPES for probing the partially-filled strongly correlated orbital symmetry of the Pr compounds in cubic symmetry, as established for the Yb systems.\cite{kanai,mori}

\begin{acknowledgments}
We thank S. Kunii for providing the single crystals of PrB$_6$ with excellent quality. We are grateful to K. Kuga, H. Aratani, Y. Aoyama, T. Hattori, M. Kawada, S. Takano, M. Murata, and C. Morimoto for supporting the experiments. AS thanks Y. Saitoh for fruitful discussions. K. T. Matsumoto, Y. Nakatani and Y. Kanai were supported by the JSPS Research Fellowships for Young Scientists. This work was financially supported by a Grant-in-Aid for Scientific Research of ``J-Physics'' (JP15H05886, JP16H01074) from MEXT, Japan and that for Scientific Research (JP16H04014) from JSPS, Japan.
\end{acknowledgments}

\end{document}